\documentstyle[epsf]{elsart}
\begin{document}
\begin{frontmatter}
\noindent Physica A {\bf 254}, 135 (1998)\\

\title{Unstable Growth and Coarsening in Molecular-Beam Epitaxy
}
\author{Lei-Han Tang}
\address{Department of Physics and Center for Nonlinear Studies\\
Hong Kong Baptist University, Kowloon Tong, Hong Kong
}

\begin{abstract}
The coarsening dynamics of three-dimensional islands 
on a growing film is discussed. It is assumed that the
origin of the initial instability of a planar surface is
the Ehrlich-Schwoebel step-edge barrier for adatom diffusion. 
Two mechanisms of coarsening are identified: (i) surface diffusion
driven by an uneven distribution of bonding energies, and
(ii) mound coalescence driven by random deposition.
Semiquantitative estimates of the coarsening time are given in each
case. When the surface slope saturates, an asymptotic
dynamical exponent $z=4$ is obtained.
\end{abstract}
\end{frontmatter}

\section{Introduction}

Nanometer-scale structures have been at the forefront of materials-related
physics research for quite some time by now. This development 
is not only driven by the intriguing electronic and optical properties 
exhibited by these systems at reduced dimensions, but also by the 
arrival of modern surface characterisation techniques which enable one 
to monitor the {\it in situ} growth and to perform morphological and 
structural analysis down to atomic dimension on a routine basis. 
The vast amount of experimental research in
this area offers ample opportunity for theoretical work in terms
of materials modelling, simulation, and analysis. 

One particular topic, which is the focus of this article,
is the appearance of nanometer to micrometer scale surface 
modulations/islands in vapor-phase epitaxy. This phenomenon is
usually regarded as the result of a growth instability, where a
planar growth front (as in the layer-by-layer growth mode)
becomes unstable. From a technological point of view, 
the instability is often undesirable in the fabrication of quantum
heterostructures (such as semiconductor superlattices),
although suggestions have been made to utilise the instability to grow
nano-scale quantum dots\cite{phys_today}.

In many epitaxial systems, a slight lattice mismatch (a few percent)
is present between the substrate and the film, usually made of a 
different kind of substance.
The strain energy builds up as the film grows thicker, and this can be
the origin of a growth instability. In fact, it has been found that
the strain in the film can relax in at least two different ways,
(i) by misfit dislocations parallel to the film surface,
and (ii) by surface rippling and ``coherent'' islanding.
Detailed calculations have been performed to show that the
gain in strain energy is enough to overcome either the dislocation
energy or the capillary energy associated with surface ripples
when the film becomes sufficiently thick. The kinetics of
the relaxation process, sometimes involving an energy barrier,
is a topic of current interest.

There is a second mechanism for unstable growth 
which can operate even in homoepitaxy.
This was first pointed out by Villain\cite{villain} in 1991 and is due to a
kinetic effect discovered by Ehrlich and Schwoebel [ES]\cite{es}. 
As seen from Fig. 1, the origin of the instability in this case 
is fairly obvious. At temperatures typical for vapor phase
epitaxy (from room temperature to several hundred degrees centigrade)
and especially for semiconductor films,
the thermal energy $k_BT$ is much smaller than the barrier energy
between neighboring surface sites. Hence adatoms move on the surface
through thermally activated hopping, with a hopping rate given by
the Arrhenius law 
\begin{equation}
r\simeq \omega_D\exp(-E_B/k_BT),
\label{Arr-law}
\end{equation}
where $\omega_D$ is the Debye frequency of the solid film.
For some surfaces, the barrier energy $E_B$ assumes a higher value
at the step edge [Fig. 1(a)]. Due to the exponential dependence
of $r$ on $E_B$, a change of $E_B$ by, say $0.5$ eV can lead to a change
of $r$ by several orders of magnitude. Thus a higher barrier at the
step edge (known as the ES barrier) can significantly reduce 
interlayer mass transport. During growth, adatoms are created at 
a constant rate by the beam.
Although individual adatoms quickly thermalise with the substrate,
the density of adatoms can be much higher than what it would be for an
equilibrium surface. When the ES barrier is present, an asymmetry in
the diffusion behaviour of adatoms around a step edge appears, as
illustrated in Fig. 1(b). Adatoms which land on the upper terrace
next to a step have nearly no access to the favorite bonding site
at the step, while those on the lower terrace can diffuse freely
to the bonding sites. A net current is thus created towards the step
from the lower side. When a gentle surface slope is present, surface
mass flow takes place in the direction of the slope. This flow is
destabilizing for a high symmetry surface
as an initially small fluctuation away from a flat
surface is amplified by the up-hill current\cite{villain}.

\begin{figure}
\epsfxsize=12truecm
\epsfbox{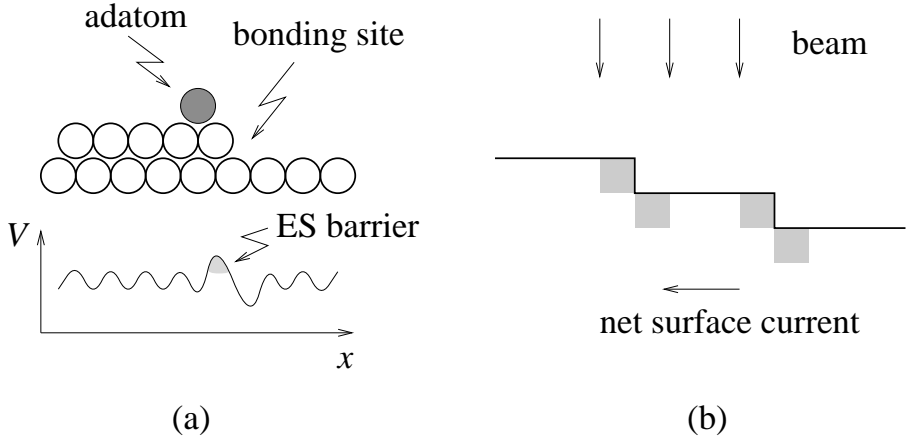}
\smallskip
\caption{(a) The Ehrlich-Schwoebel (ES) barrier prevents
a thermalised adatom to jump down a step to reach the
bonding site. (b) When a gentle surface slope is present,
atoms arriving in the shaded area experience a diffusion bias
due to the ES barrier: those landing on the lower side of a
step are attracted by the step while those above are repelled.
This asymmetry leads to a net surface diffusion current
in the direction of the surface slope.}
\end{figure}

A number of kinetic Monte Carlo studies of solid-on-solid models
of growth have been carried out. When the ES barrier is incorporated
in the adatom hopping rates, three-dimensional surface modulations
are indeed observed\cite{johnson,stroscio,siegert,pavel,family}.
The evolution of the surface modulation consists of two stages,
analogous to the spinodal-decomposition phenomena in
binary fluids and magnetic systems\cite{bray}. In the first stage, 
mounds with a characteristic lateral size grow rapidly in their
height until their slope approaches a saturated value.
This is known as the {\it initial transient regime}.
(In some models, steepening continues to occur beyond the transient
regime, but at a much slower rate.)
In the second stage, the typical lateral size $L$ of the mounds
grows with continuing deposition. This phenomenon is known as 
{\it coarsening}.
Both simulations and some experiments\cite{stroscio,thurmer} have shown that 
the dependence of $L$ on the film thickness $H$ (see Fig. 2)
can be fitted to a power-law
\begin{equation}
L\sim H^{1/z}.
\label{size-scaling}
\end{equation}
The exponent $1/z$ generally lies in the range $0.15$-$0.25$.
The typical height of the mounds $W$
can also be fitted to a power-law,
\begin{equation}
W\sim H^\beta,
\label{mound-height}
\end{equation}
where the exponent $\beta$ can be as small as $0.25$, or as big as $0.5$.

\begin{figure}
\epsfxsize=8truecm
\hbox{\hskip 3truecm\epsfbox{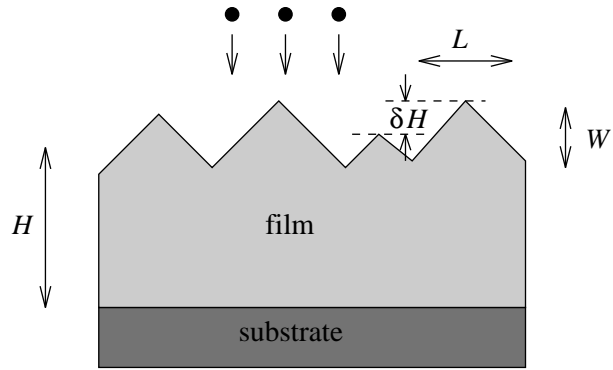}}
\smallskip
\caption{Schematic illustration of the mound morphology
during epitaxial growth. The typical mound size $L$ 
increases with the film thickness $H$.}
\end{figure}

The origin for the apparent nonuniversal value of the scaling exponents
is unclear. This is in fact not surprising in view of the fact there
is no clear picture why coarsening should take place in the first place.
Due to this lack of understanding, there is quite a bit of uncertainty
with regard to the way experimental or simulational data should be
analysed. In the following we report some initial attempts
in addressing these issues\cite{tsv97}.

\section{Bonding energy driven coarsening: relation to magnetic systems}

Coarsening laws of the type (\ref{size-scaling}) are quite familiar in
statistical mechanics. In an Ising magnet, quenching a system
from the high temperature disordered phase to below $T_c$, 
ordered domains form locally and then coarsen with time (see Fig. 3).
The characteristic length $L$ of the domain structure increases with
time $t$ as $L\sim t^{1/z}$. The value of the exponent $z$ depends on 
whether the dynamics is conservative (e.g., as in Kawasaki
spin exchange dynamics) or otherwise\cite{bray}.

Several groups\cite{krug} have made the suggestion that, treating the gradient 
of the surface ${\bf m}=\nabla h$ as an order parameter, there is a 
direct analogy between the growth problem and the ordering dynamics
of a planar magnet in two dimensions. So far, this analogy has
been explored only formally and the similarities and differences between the
two classes of problems have not been pursued in detail.

When the deposition noise in the growth problem can be ignored,
there is indeed a close analogy between the growth problem and
the Ising problem under conserved dynamics (Model B), though the coarsening
exponent is different in two dimensions due to a difference
in the transport mechanism operative. To see this, let us first
review how the coarsening law in the Ising case under conserved
dynamics is obtained.
From Fig. 3, we see that the domain structure is fairly convoluted.
The driving mechanism for coarsening is the surface tension
associated with the domain walls. To reduce this excess free 
energy, spins need to diffuse around in such a way so as to eliminate 
points on the interface with the highest curvature.
When the typical size of the pattern is $L$, the typical
difference in chemical potential between favorable and unfavorable
places along the interface is given by $\Delta\mu\sim L^{-1}$.
At moderate temperatures and sufficiently large $L$, 
the dominant mechanism for transport is bulk diffusion,
as anyone who has watched the evolution of the pattern on
a computer can contest. Thus on average we expect a bulk
current $j_B\sim \Delta\mu/L\sim L^{-2}$ across the domains and
a rate of transport $r\simeq L j_B\sim L^{-1}$ over an
area $L^2$. For the coarsening to take place, a finite
fraction of spins in such an area need to diffuse through,
yielding a coarsening time
\begin{equation}
\tau\simeq L^2/r\sim L^3.
\label{ising}
\end{equation}
This is the famous Lifshitz law.

\begin{figure}
\epsfxsize=8truecm
\hbox{\hskip 3truecm\epsfbox{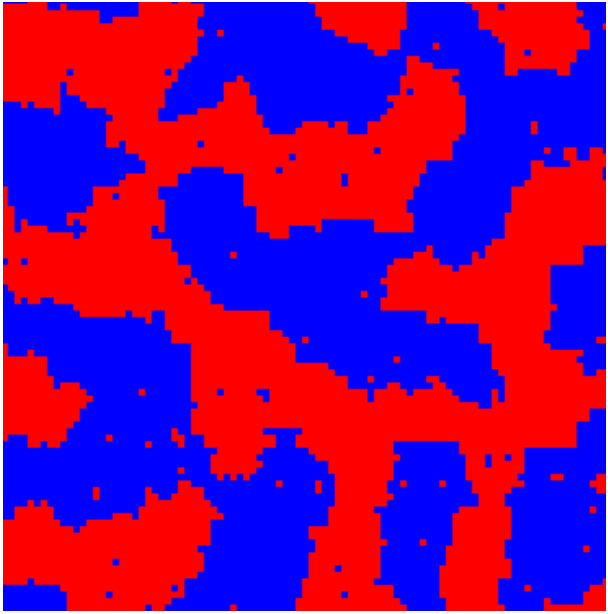}}
\smallskip
\caption{Domain structure in an Ising model after quenching
from the disordered phase to a temperature $T=0.7T_c$.
The simulation uses Kawasaki spin exchange dynamics.
Note the isolated spins which diffuse through domains
of opposite magnetisation. This bulk diffusion is responsible for
the Lifshitz coarsening law $z=3$.}
\end{figure}

Let us now see how the above picture can be modified to
explain coarsening in the growth problem.
For simplicity, let us first consider the mass transport
between two neighboring mounds in a typical late-stage situation,
as depicted in Fig. 4. 
The central part of each mound consists of roughly concentric rings of steps.
The two mounds are joined by a ``ridge terrace''. The outer rim
of the ridge terrace has convex parts on either side
and concave parts in the middle. 

We assume that the newly deposited atoms are quickly captured
by their nearby steps. These immediate bonding sites, however,
may not be the energetically most favorable sites on the entire
surface. This can be easily seen from Fig. 4, where 
sites on the convex parts of a step
on average offer less lateral bonding than those on the concave
parts. This is very similar to the surface tension
effect in the domain coarsening problem discussed above.
One can thus assign a chemical potential difference $\Delta\mu$ between
the convex and concave parts. Assuming the steps are locally
thermal equilibrated, we have
\begin{equation}
\Delta\mu\simeq \gamma_s/L,
\label{delta-mu}
\end{equation}
where $\gamma_s$ is the step free energy per unit length,
and $L$ is the typical lateral scale in question
(e.g., the distance between two centers).

\begin{figure}
\epsfxsize=8truecm
\hbox{\hskip 3truecm\epsfbox{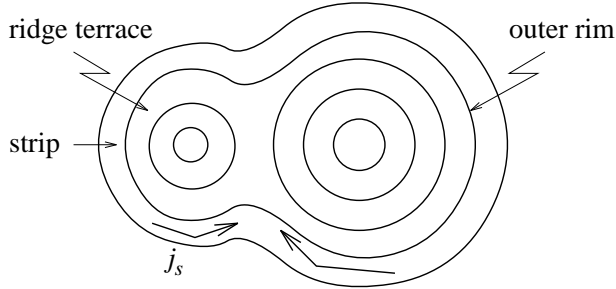}}
\smallskip

\caption{Top view of two neighboring mounds of unequal size.
Better lateral bonding for surface atoms is achieved at the
concave parts of the closed steps.
This mechanism results in an inward mass current $j_s$.
}
\end{figure}

Although the dependence of $\Delta\mu$ on $L$ is the same
as in the magnetic case, the transport mechanism
here is different. This has to do with the geometry of the
problem. As we have mentioned in the previous section,
mounds are there because of the ES barrier, which suppresses 
interlayer transport. 
In order for an atom on the outer rim to reach the more
favorable bonding sites at the center, it must diffuse
either along the step, or on the terrace next to the step.
In both cases, the diffusion is effectively one-dimensional
as compared to the bulk diffusion in the Ising case.
This gives rise to a slower coarsening law.

A semi-quantitative estimate of the coarsening time
can now be obtained as follows.
The average mass current on the ridge terrace due to the
chemical potential difference (\ref{delta-mu}) is given by,
\begin{equation}
j_s\simeq D_s\Delta\mu/L,
\label{j_s}
\end{equation}
where $D_s$ is a kinetic transport coefficient.
(Note that $D_s$ may depend on the mound slope when 
terrace diffusion dominates over ledge diffusion.)
The same mechanism is expected to operate also in other layers below
the ridge terrace, though $j_s$ decreases due to the decreasing curvature
of the steps.
Let $Q$ be the number of layers with significant inward mass transport,
the total inward mass current is given by,
\begin{equation}
J_s\simeq Qj_s\simeq D_s\Delta\mu{Q\over L}.
\label{total-current}
\end{equation}

The process described above yields a gradual outward expansion
of the neck region connecting the two mounds.
When the amount of mass $M$ transported reaches a value comparable
to the volume needed to fill in the gap, $WL^2$, the two mounds
coalesce. This leads to an estimate of the coarsening time,
\begin{equation}
\tau_b=M/J_s\simeq L^4{W\over \gamma_sD_s Q}.
\label{tau-b}
\end{equation}
With $Q\simeq W$, Eq. (\ref{tau-b}) yields $z=4$.

\section{Noise driven coarsening}

Coarsening driven by an uneven distribution of
bonding energies on the surface is effective only when
the diffusion constant $D_s$ is sufficiently large.
Surface transport is strongly influenced by the substrate
temperature. Thus, when the substrate temperature
is low, coarsening due to this
mechanism is extremely slow. In this section we discuss
another coarsening mechanism: coarsening due to random fluctuations.
Such a mechanism is present at all temperatures
and it dominates when the previous mechanism
is rendered ineffective by a low substrate temperature.

To illustrate how the noise mechanism works, let us
consider a simple model where the inter-mound
mass transport can be ignored completely.
The height of each mound is now solely determined
by the flux it receives from the beam.
The total number $N$ of atoms arriving
within a surface area $L^d$ when the film grows to a thickness $H$ is
$N\simeq HL^d/\Omega$. Here $d$ is the dimension of the surface,
and $\Omega$ the volume occupied by a single atom in the grown film.
Due to the random nature of the vapor phase epitaxy,
this number has a fluctuation
\begin{equation}
\delta N\simeq N^{1/2}\simeq H^{1/2}L^{d/2}\Omega^{-1/2}.
\label{delta-N}
\end{equation}

From the geometry illustrated in Fig. 2, we see that the height
difference between neighboring mounds can not exceed significantly
the typical height corrugation $W$ of the mound array.
This imposes a coarsening condition: the volume fluctuation
$\Omega\delta N$ of a mound can not exceed the volume itself,
$WL^d$. Equating the two terms, we obtain a coarsening time 
(measured in film thickness)
\begin{equation}
H\simeq W^2L^d/\Omega.
\label{noise-scaling}
\end{equation}

In the regime where the power-laws (\ref{size-scaling})
and (\ref{mound-height}) are well obeyed, Eq. (\ref{noise-scaling}) yields,
\begin{equation}
{2\beta\over d}+{1\over z}={1\over d}.
\label{scaling-law}
\end{equation}
If the mound slope $s\simeq W/L$ saturates to a constant,
the above equation gives,
\begin{equation}
\beta={1\over z}={1\over d+2}.
\label{beta}
\end{equation}

A few comments on the coarsening laws derived here are in order.
The noise mechanism does not explain why mounds form in the
first place, nor does it tell us anything about the mound shape
and slope. However, based on the assumption that there is an
underlying mechanism for the formation of stable mounds,
and that inter-mound transport is sufficiently slow, a
relation between the geometrical parameters $L,W$ and $H$
characterising the mound array follows.
This relation, of course, does not involve any of the
microscopic details of the surface dynamics, provided
the assumptions are met.

One might wonder why noise-fluctuation does not generate
mound splitting. In simulation studies, mound splitting
has indeed been observed, but it happens with a much smaller
probability than the reverse process of mound coalescence.
This obviously has to do with the assumed stability of
individual mounds. 

\section{Which mechanism wins?}

From Eq. (\ref{beta}), we see that the dynamical exponent
$z$ due to noise mechanism is equal to 
\begin{equation}
z_n=2+d,
\label{z-noise}
\end{equation}
asymptotically at large $L$. 
If one takes the point of view that the bonding energy
driven coarsening always invokes some sort of transport
in $d-1$ dimensions, then its corresponding dynamic exponent
is given by
\begin{equation}
z_b=4,
\label{z-energetic}
\end{equation}
in all dimensions.
Thus for $d<2$ the noise mechanism dominates, while for
$d>2$ the bonding energy mechanism dominates, with $d=2$
being the marginal case!

For $d=2$, the bonding energy
mechanism is expected to dominate at high substrate temperatures,
while the noise mechanism should dominate at low temperatures.
A quantitative criterion can be constructed by comparing the
timescales involved in each case.
At a given deposition rate $F$,
Eq. (\ref{noise-scaling}) yields a coarsening time 
$\tau_n=H/F=W^2L^2/(\Omega F)$ on scale $L$ under the noise mechanism. 
Comparing it with $\tau_b$ given by (\ref{tau-b}),
we obtain a dimensionless parameter
\begin{equation}
\alpha={\tau_b\over \tau_n}={\Omega F\over\gamma_sD_s s^2},
\label{alpha}
\end{equation}
where $s=W/L$ is the saturated slope of the mounds.
The bonding energy mechanism dominates when $\alpha\ll 1$ ($\tau_b\ll\tau_n$),
while the noise mechanism dominates when $\alpha\gg 1$ ($\tau_b\gg\tau_n$).

The next question is whether we can distinguish the two mechanisms
based on a measurement of the surface profile.
This turns out to be possible because the noise mechanism gives
a specific relationship between the geometrical parameters
$H, W$ and $L$, all can be measured directly when we know the
surface profile and film thickness. The quantities $W$ and $L$
can be defined precisely by examining the two-point correlation
function\cite{stroscio}
\begin{equation}
G({\bf x},t)=\langle \tilde h({\bf x}_0,t)
\tilde h({\bf x}_0+{\bf x},t)\rangle,
\label{two-point}
\end{equation}
where $\tilde h=h-H$ is the fluctuation away from a flat film surface.
We can identify the mound height $W$ with the root-mean-square width
of the corrugated surface,
\begin{equation}
W=\langle\tilde h^2\rangle^{1/2}=G^{1/2}(0,t),
\label{rms-width}
\end{equation}
and the mound size $L$ with the first zero of $G$, either along
a given crystallographic direction or its radial average.
(The function $G$ typically has many oscillations for a mounded
surface because of the anticorrelation of mound slopes over a distance $L$.)
From the three geometrical quantities, 
we introduce a dimensionless parameter,
\begin{equation}
R={\Omega^{1/2}H^{1/2}\over WL^{d/2}}.
\label{ratio}
\end{equation}
If coarsening is driven by noise, $R$ should be of order 1.
Otherwise, $R$ should depend on the details of the surface
dynamics, but its value should never exceed the limit set
by the noise mechanism.
(It is easy to see that $R$ also measures the ratio between the
excess material in a given mound due to fluctuations in the
deposition rate, $\delta N$, and the total mound volume $WL^d$.)

An interesting property of (\ref{ratio}) is that, for $d=2$,
$R=$const implies $z=4$ scaling of coarsening time
when the mound slope $s=W/L$ saturates.
Thus the constancy of $R$ can also be used as a test
for the validity of such a scaling law. It should be
noted that, even when this scaling does not hold,
the value of $R$ can still be used to distinguish between
noise driven ($R\simeq 1$) and bonding energy driven 
($R\ll 1$) coarsening.

In the real MBE experiment, both mechanisms are present
and they play somewhat complementary role in the coarsening
process. For example, the bonding energy mechanism is totally
ineffective if we have a perfectly regular array of mounds.
Symmetry then tells us that no mounds can grow at the expense
of other mounds, and the whole pattern should stay put
(although the shape of individual mounds may evolve).
However, deposition noise will generate fluctuations
in mound size and, coupled to the surface transport,
eventually leads to coarsening.
The surface transport is expected to become more significant
during the final stage of a coalescence event, as there
the curvature effects become greater, and the diffusion
length becomes smaller. (If diffusion is totally ineffective,
we would have the random deposition model with no mounds
whatsoever. Thus the very existence of mounds demands some
form of surface diffusion be present, though the diffusion
may not be efficient enough to account for coarsening.)

Based on the above discussion, Tang, Smilauer and Vvedensky\cite{tsv97}
proposed the following picture of mound coarsening:
In the late-stage coarsening regime where mounds have acquired
their quasi-stationary shape, deposition noise is responsible
for generating the height difference (or equivalently, size disparity) 
between mounds through random fluctuations.
Geometry sets a maximum $\delta H_m=W$ for the possible height
difference. Bonding energy driven transport, on the other hand, speeds
up the coarsening process by eliminating mounds which become
too small compared to the rest. Effectively, it makes the surviving
mounds more uniform in size. The quantitative importance of this effect
depends on the parameter $\alpha$.

\section{Comparison with simulation and experimental results}

The mound coarsening problem has been examined in quite a few
numerical and experimental investigations, although a consensus
on the coarsening law is yet to emerge. From the discussion
reported above, it is seen that the coarsening time depends
on a number of factors whose role can be different on different
length scales and for different systems, experimental or otherwise. 
Therefore a careful assessment of 
the pattern and efficiency of surface mass
transport, the importance of noise fluctuations, etc., is needed
before a meaningful comparison of the values for $z$, 
often extracted from a limited range of data, can be made.
This unfortunately has not been done for most of the
experimental and simulational results that have been reported,
making it difficult to test our theory. In the following
we mention two exceptions: coarsening in $d=1$, and
a model studied by Smilauer and Vvedensky.

In the case $d=1$, the noise driven coarsening picture is consistent 
with past simulation work, particularly the one by 
Kawakatsu and Munakata.\cite{km} 
The model they studied can be interpreted in the surface
growth context\cite{krug}.
Numerical studies yields $z=3$ which is consistent with
(\ref{z-noise}).

In the second example, we mention a two-dimensional solid-on-solid model 
introduced by Smilauer and Vvedensky to model growth of both 
metal and semiconductor films with an ES barrier.\cite{pavel}. 
The hopping rate of a surface atom follows the Arrhenius law
(\ref{Arr-law}), with the barrier energy $E_B$ depending on
the local bonding configuration, and the ES barrier when the move
involves jumping down a step. Simulations were carried out at various
temperatures for two sets of surface energy parameters.\cite{pavel}
In all cases examined, a mounded surface appears after an
initial transient and the structure is observed to coarsen
in time. In a recent work, Tang, Smilauer and Vvedensky computed the 
parameter $R$ using the simulation data as a function of
layer thickness $H$. They observed that, in all cases,
$R$ saturates to a constant value at late times, suggesting
that the dynamic exponent $z$ is indeed four. 
The actual asymptotic value of $R$ depends on the surface
temperature, being close to 1 at low temperatures, and
much smaller than 1 at high temperatures. It is also
found that the transition between the two regimes is
fairly abrupt as the substrate temperature is increased,
reflecting the exponential dependence of surface transport
on temperature as suggested by the Arrhenius law.
All these conclusions are consistent with the picture
developed above.

To provide a more stringent test of the theory, future work
should attempt to compute the step tension parameter $\gamma_s$
and the diffusion constant $D_s$, either directly or 
in a numerical experiment with a suitable initial configuration
of the surface. Much numerical work needs to be done
in this direction.

\section {Conclusions}

The two mechanisms discussed in this paper, bonding energies and
noise, all lead to coarsening of the mound array. Based on
the pictures developed, quantitative estimates for the coarsening
time are obtained. These estimates in particular allows one to
derive asymptotic scaling laws when the mound slope saturates.
In two dimensions, one obtains $z=4$. It has been shown that
these estimates are consistent with some existing simulational
and experimental work. We believe the ideas and
methods presented here offer a systematic way to analyse future
simulational and experimental results.

{\bf Acknowledgements:} The results reported here grew out of
a collaboration with P. Smilauer and D. Vvedensky.
I would like to thank the conference organisers for
a very enjoyable and stimulating meeting.

\end{document}